\documentclass[journal,12pt,onecolumn]{IEEEtran}

\usepackage[ruled]{algorithm2e}

\SetAlFnt{\small}
\SetAlCapFnt{\small}
\SetAlCapNameFnt{\small}
\SetAlCapHSkip{0pt}
\IncMargin{-\parindent}

\usepackage{moreverb,algorithmic,subfigure}
\usepackage{amsmath}
\usepackage{amssymb,lscape}
\usepackage{tabularx}
\usepackage{balance}

\newcommand{\be}{\begin{equation}}
\newcommand{\ee}{\end{equation}}

\usepackage{cite}
\usepackage{epsfig}
\usepackage{times}
\usepackage{graphicx}
\usepackage[figuresright]{rotating}
\usepackage{subfigure}
\usepackage{amssymb}
\usepackage{amsmath}
\usepackage{setspace}
\usepackage{rotating}
\usepackage{multirow}
\usepackage{wrapfig}
\usepackage{booktabs}
\usepackage{array}
\usepackage{comment}
\usepackage{subfigure}
\usepackage{epstopdf}
\usepackage{epsfig}
\usepackage{amsmath}
\usepackage{amssymb}
\usepackage{times}
\usepackage{subfig}

\usepackage[ruled]{algorithm2e}

\SetAlFnt{\small}
\SetAlCapFnt{\small}
\SetAlCapNameFnt{\small}
\SetAlCapHSkip{0pt}
\IncMargin{-\parindent}

\usepackage{moreverb,algorithmic,subfigure}
\usepackage{amsmath}
\usepackage{amssymb,lscape}
\usepackage{tabularx}
\usepackage{times}
\usepackage{float}

\usepackage{caption}
\usepackage{varwidth,xcolor}

\usepackage{colortbl}

\begin{document}

\title{\huge{Transform-Domain Classification of Human Cells \\ based on DNA Methylation Datasets}}


\author{Xueyuan~Zhao and Dario~Pompili
\thanks{\IEEEcompsocthanksitem Xueyuan Zhao and Dario Pompili are with the Department of Electrical and Computer Engineering, Rutgers University--New Brunswick, NJ, USA. E-mails: \{xueyuan.zhao, pompili\}@rutgers.edu \protect\\ \protect
}
}


\clearpage\maketitle
\thispagestyle{empty}
\doublespacing

\begin{abstract}

A novel method to classify human cells is presented in this work based on the transform-domain method on DNA methylation data. DNA methylation profile variations are observed in human cells with the progression of disease stages, and the proposal is based on this DNA methylation variation to classify normal and disease cells including cancer cells. The cancer cell types investigated in this work cover hepatocellular (sample size $n$ = 40), colorectal ($n$ = 44), lung ($n$ = 70) and endometrial ($n$ = 87) cancer cells. A new pipeline is proposed integrating the DNA methylation intensity measurements on all the CpG islands by the transformation of Walsh-Hadamard Transform~(WHT). The study reveals the three-step properties of the DNA methylation transform-domain data and the step values of association with the cell status. Further assessments have been carried out on the proposed machine learning pipeline to perform classification of the normal and cancer tissue cells. A number of machine learning classifiers are compared for whole sequence and WHT sequence classification based on public Whole-Genome Bisulfite Sequencing~(WGBS) DNA methylation datasets. The WHT-based method can speed up the computation time by more than one order of magnitude compared with whole original sequence classification, while maintaining comparable classification accuracy by the selected machine learning classifiers. The proposed method has broad applications in expedited disease and normal human cell classifications by the epigenome and genome datasets.

\end{abstract}

\section{\bf{Background}}\label{sec:intro}

\begin{figure} \label{chromatin}
\centering 
\includegraphics[width=0.66\textwidth]{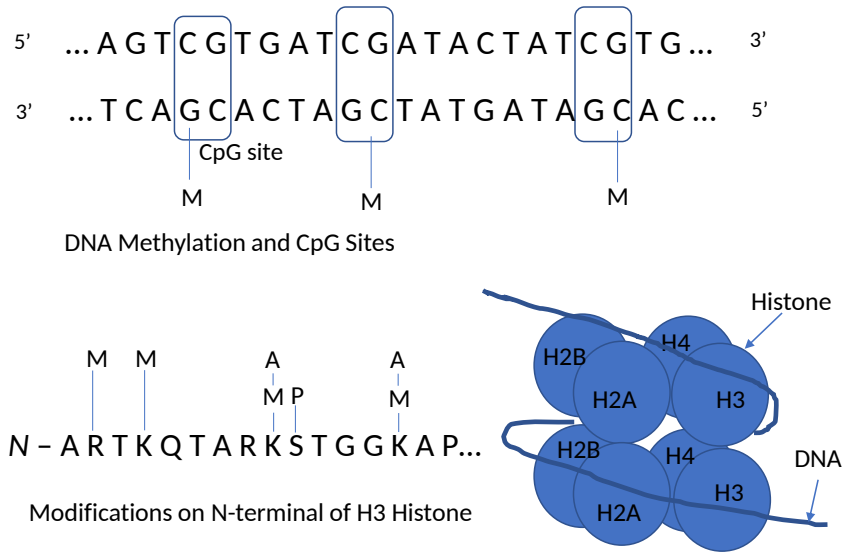}
\caption{Chromatin modifications in epigenome: 1)~DNA methylation: DNA sequence with cytosine followed by guanine from 5' to 3', modified by methyl (denoted by M); 2)~Histone modification: histone composed of H2A, H2B, H3, and H4, modified by methyl, acetyl (denoted by A) and phosphate (denoted by P).}\label{fig:chromatin}
\end{figure}

{
DNA methylation is one of the major epigenetic effects and can be measured by Whole-Genome Bisulfite Sequencing~(WGBS). DNA methylation is defined by the biological phenomena that methyl is attached to certain regions of the genome. In the DNA sequence, there are particular regions, called \emph{CpG sites}, which are defined as a cytosine followed by a guanine in the 5' $\rightarrow$ 3' direction in the DNA sequence, as depicted in Fig.~\ref{fig:chromatin}. The DNA methylation raw measurement contains the methylation intensities at CpG sites of the whole genome. The DNA methylation frequently occurs at genome areas that are rich in CpG sites, named \emph{CpG islands}. The CpG island is a cluster of CpG sites in close proximity usually located in gene-coding regions. The DNA methylation effect is highly correlated with CpG islands. The DNA methylation is associated with the type and status of the cells and is varying due to various biological factors. For these reasons, studying the DNA methylation data at CpG island is of particular significance to identify the status of the cells. DNA methylation WGBS datasets are one major epigenome dataset applied to the current study. The histone modification is another important epigenetic effect in addition to DNA methylation. The histone is a protein structure surrounded by DNA sequence and is composed of four proteins, named H2A, H2B, H3, and H4. Each composing protein can be expanded to an N-terminal-tailed amino-acid sequence, as denoted in Fig.~\ref{fig:chromatin}. The histone modification is the effect that certain chemical structures are bonded to these amino-acid sequences. The histone can be modified by methyl, acetyl, and phosphate. In addition to DNA methylation and histone modification, the DNA accessibility and RNA-seq are also the information from epigenome. the DNA accessibility regions are the open genome regions of regulatory functions that are sequenced by DNase-seq/ATAC-seq. The RNA-seq contains the information of RNA splicing, gene expression, and post-transcriptional modifications. The DNA methylation, histone modification, DNA accessibility, and RNA-seq information jointly determine the epigenome of the cells. Due to the variation of DNA methylation between disease and non-disease cells, the DNA methylation can be a key indicator of the state of cells for the classification of the cell types and disease conditions.
}

\textbf{Existing work:}
{
On the disease association study, it has been discovered that the DNA methylation variations are associated with the disease of cancer. Aberrant patterns of DNA methylation in cancer cells are found in a number of research works. It is found that the DNA methylation profile of Small Cell Lung Cancer~(SCLC) samples~\cite{Poirier15} has distinct patterns. The DNA methylation pattern is studied on Renal Cell Carcinoma~(RCC)~\cite{Slater13}, where the differentially-methylated genetic locations are identified for this type of cancer. The genetic location-specific DNA methylation changes have been studied for anaplastic large-cell lymphoma~(ALCL)~\cite{Hassler16}. The co-dependency of DNA and DNA methylation during the brain tumor evolution has been studied in~\cite{Mazor15}. The genes and DNA methylation variation that have associations with endometrial cancer~\cite{Farkas16} and lung cancer~\cite{BjaanA16} have also been investigated. In addition, regarding cancer related classification studies, the DNA methylation datasets are analyzed with RNA sequencing data~\cite{Cappelli18} and genome data~\cite{Weitschek18} for classification of cancer cells and information extraction from cancer datasets.
}

DNA methylation profiles of neural degenerative disease cells also show aberrant patterns compared to normal cells. The CpG regions associated with genetic loci of Alzheimer's disease are validated~\cite{DeJager14} and differentially methylated regions are identified. Brain DNA methylation in selected genetic loci was found~\cite{Yu15} to be associated with the pathology of Alzheimer's disease. It is found that global DNA methylation is increased in a study involving late-onset Alzheimer's disease group~\cite{Francesco15} and the healthy control group. On schizophrenia disease, by the genome-wide DNA methylation analysis~\cite{Wockner14}, unique genetic locations are found to be associated with schizophrenia. Differential expression and methylation of genes~\cite{Chen14} are found for schizophrenia and bipolar disorder patients. There are more recent works discussing the genetic associations of DNA methylation in the schizophrenia disease~\cite{Deng18,McKinney18,Jiang17,Huda17}. The existing works on DNA methylation disease association are focusing on specific CpG islands and genetic locations where the disease cells have aberrant patterns compared with non-disease cells.

{
Existing cell-classification works based on DNA methylation are focusing on identifying certain CpG islands with most significant changes of DNA methylation between the disease and non-disease cells, and on the association study of the DNA methylation on these CpG islands to the disease status. A classification method is proposed to identify the cancer-driver genes~\cite{Celli18} that have significant relevance in terms of statistical DNA methylation variation between cancer and non-cancer disease cells. Note that, our proposal is different from these existing works in the sense that it does not need to identify genes that have disease associations. The DNA methylation vector is directly transformed by WHT and the resulting sequence is the input to the machine learning classifiers for the disease cell classification. The classification is studied based on the DNA methylation of $22$ selected CpG islands~\cite{Langevin15} for cancer cell classification. The classification identifier of CpG Island Methylator Phenotype~(CIMP)~\cite{Borssen16} is applied to classify the patients with T-Cell Acute Lymphoblastic Leukemia. A defined value, named \emph{FR-pair}, is proposed~\cite{Li15} to classify the normal and cancer cells based on the DNA methylation beta value vector. To the best of our knowledge, no transform-domain method has been proposed on the DNA methylation vector for cell classification by other research groups.
}

 \begin{table}
 \centering
 \caption{The GEO datasets in the data analysis of this work.} \label{table_datasets}
 \begin{tabular}{|p{2.3cm}|p{1.6cm}|p{1.6cm}|p{1.6cm}|p{4.3cm}|}
 	\hline
 	\textbf{Dataset}& \textbf{Total Samples}& \textbf{Disease Samples}& \textbf{Normal Samples}&\textbf{Disease} \textbf{(type of cancer)}\\
 		\hline
     GSE17648&44&22&22&colorectal cancer
 \\
     \hline
     GSE73003&40&20&20&hepatocellular cancer
 \\
       \hline
     GSE63384 &70& 35 & 35 &lung cancer
 \\
 	\hline
    GSE40032 &87&64&23&endometrial cancer\\
     \hline
   \end{tabular}
 \end{table}

Existing works on applying machine learning/deep learning approaches to DNA methylation are mostly on the regression computation of DNA methylation to predict the methylation states at unknown CpG sites. The deep convolutional neural network is designed in program CpGenie~\cite{Zeng17} for DNA methylation prediction at CpG sites, and the methylation prediction results are applied to the prediction of the impact of functional variants at non-coding regions on the DNA methylation. A hybrid deep learning network, named DeepCpG~\cite{Angermueller17}, has been proposed to obtain the regression results of DNA methylation states based on the single-cell profiled DNA and DNA methylation data, where the proposed structure is composed of a convolutional neural network to process DNA data, and a recurrent neural network to process DNA methylation data. Support Vector Machine~(SVM) and stacked denoising autodecoder have been applied~\cite{Wang16} on the prediction of binary DNA methylation states, and it is found that the stacked denoising autodecoder has close and slightly lower prediction accuracy than SVM. A gradient boosting method, named BoostMe~\cite{Zou17}, is studied on the prediction of the DNA methylation and shows performance gains over random forests and DeepCpG. Review work on deep learning on bioinformatics has also been reported in the same work. Other related computational methods studying the DNA methylation datasets include continuous-time Markov chain~\cite{Capra14} and support vector regression model~\cite{Xu15}. In these existing works, deep-learning techniques are applied to the regression to predict the DNA methylation states.


\textbf{Our contributions:}
In contrast to existing works on human disease cell classification which identify the most aberrant CpG islands at genetic locations, we are proposing new classification approach based on the discovery of the low dimension and three-step property of DNA methylation profile after transform-domain method of Walsh-Hadamard Transform (WHT). We find that the WHT of the DNA methylation beta value vector produces a low dimension transform-domain vector. This transform domain vector has values with three steps of amplitude levels. The rest of the transform-domain vector has low amplitude similar to noisy elements. This low dimension and three-step property of DNA methylation with WHT is one of the main contributions and was published in our work~\cite{ICBCB17}. In this journal extension, this dicovery is extended to the performance comparison between the proposal and the whole sequence classification by comprehensive evaluations of typical machine learning classifiers suited for this application. The comparisons are done on classification of disease and non-disease cells. To summarize, the contributions of our work are:

\begin{table*}
\centering
\caption{Step values and ranges of selected normal tissue cells in the GEO datasets.} \label{table_steps}
\begin{tabular}{|p{3cm}|p{1.3cm}|p{1.3cm}|p{1.3cm}|p{1.3cm}|p{1.3cm}|p{1.3cm}|p{1.3cm}|}
\toprule[1.4pt]
	\textbf{Dataset} & \textbf{Step 1 Mean} &\textbf{Step 2 Mean} &\textbf{Step 3 Mean} &\textbf{Step 1 Range} &\textbf{Step 2 Range} &\textbf{Step 3 Range}\\
	\hline
    GSE17648(27K) &0.0420&0.0248 &0.0083&2:4& 5:8 &9:32\\
	\hline
   GSE73003(27k) &0.0420 &	0.0254 &	 0.0085 & 2:4 &5:8 &9:32\\
	\hline
    GSE63384(27k)&0.0429 & 0.0264 	&0.0088 &2:4 &5:8 &9:32 \\
    \hline
    GSE40032(27K)&0.0412 & 	0.0248& 	 0.0083 & 2:4 &5:8 &9:32\\
    \hline
	 GSE70977(450K)&0.1030& 	0.0260& 	 0.0070 & 2:4 &5:16& 17:64\\
\bottomrule[1.4pt]
\end{tabular}
\end{table*}

\begin{itemize}

\item The low-dimension and three-step properties are found on the WHT transform-domain vector of DNA methylation beta value vector. The transform-domain vector is unique for distinct tissue cell and shows difference between disease and non-disease cells;

\item A new cell classification pipeline is proposed based on the transform-domain method of WHT. The classifications of disease and non-disease cells are tested on public datasets, and WHT transform-domain data and whole original sequence data are compared on the classification accuracy and computation time reduction.

\item It is found that, the proposal can speed up the disease cell classification by more than an order of magnitude compared with whole original sequence classification, while maintaining comparable classification accuracy by the machine learning classifiers of best performance. Therefore the proposal is well-suited for expedited classification of disease cells based on DNA methylation datasets.

\end{itemize}

\textbf{Broader applications:}
This approach is also valuable for disease risk evaluations, for example, cancer risk or neural degenerative disease risk evaluations. The DNA methylation variations are associated with several \emph{internal factors} such as the disease status, the age, the genetic information, and the tissue types. However, \emph{external factors} such as dietary and lifestyle/living conditions complicate the study of these variations, since both internal and external factors jointly determine the DNA methylation profile status, for a particular tissue cell at a particular time. For example, the tissue cells of potential cancer patients can be sequenced and subsequently classified by the proposed pipeline to obtain its disease status.

\textbf{Article outline:}
This article is organized as follows. The proposed pipeline is described in Sect.~\ref{method}. The cell classification results are given in Sect.~\ref{results}. Finally, conclusions are drawn in Sect.~\ref{conclusion}.

\begin{figure*}
\centering
\begin{varwidth}{0.5\linewidth}  
\subfigure[]{\includegraphics[width=9cm]{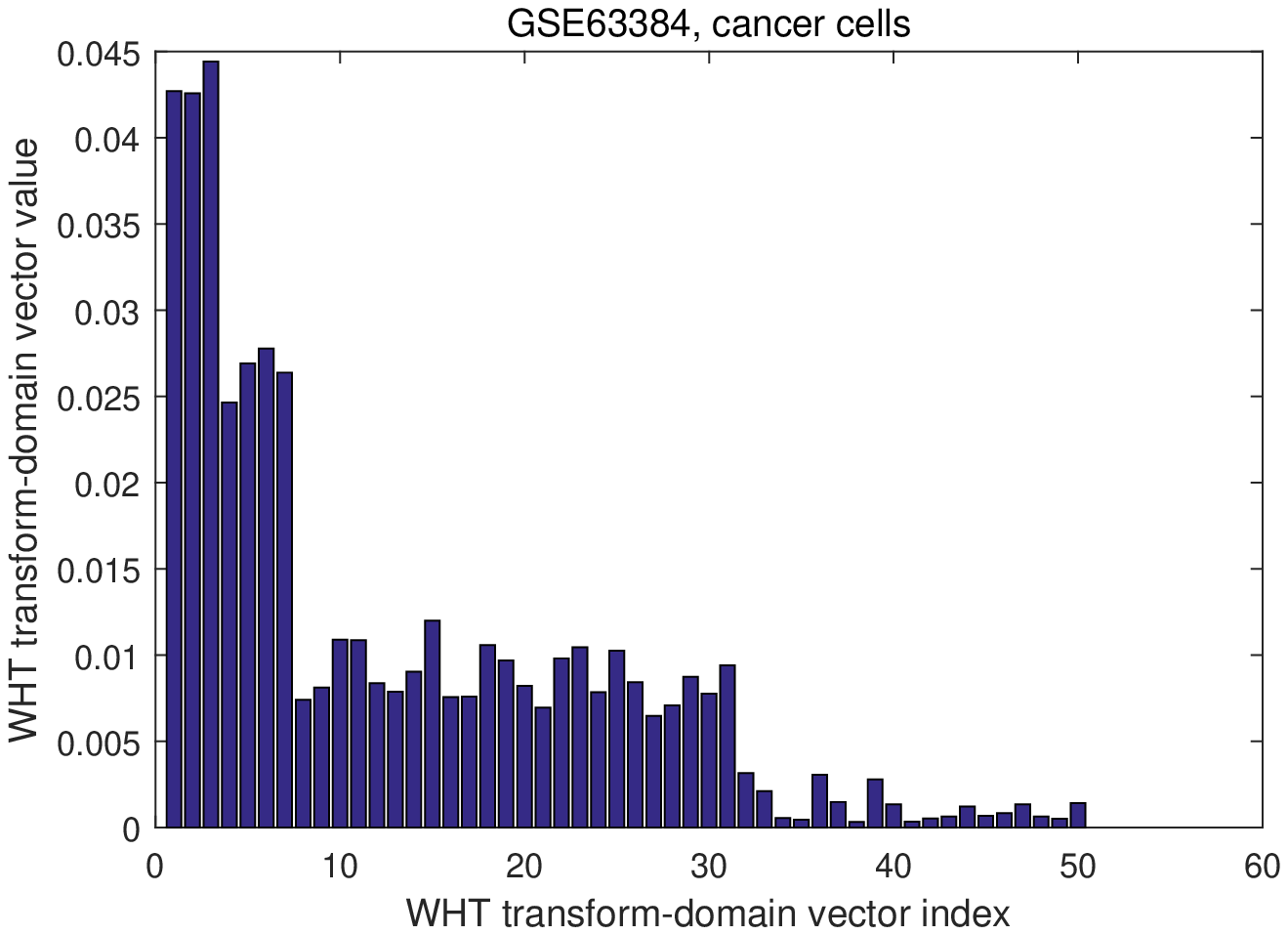}}\\
\subfigure[]{\includegraphics[width=9cm]{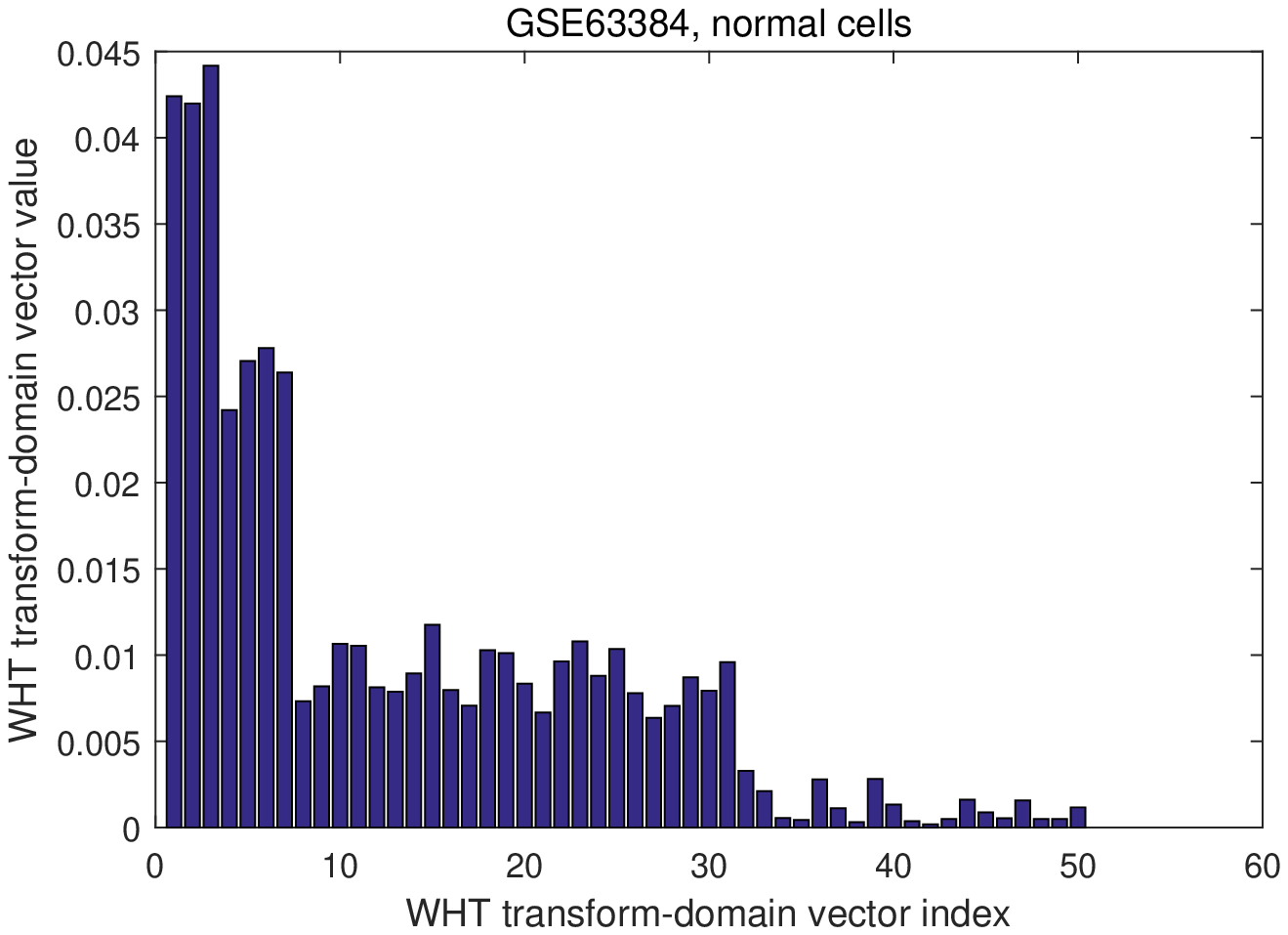}}
\end{varwidth}
\begin{varwidth}{0.5\linewidth}  
\subfigure[]{\includegraphics[width=9cm]{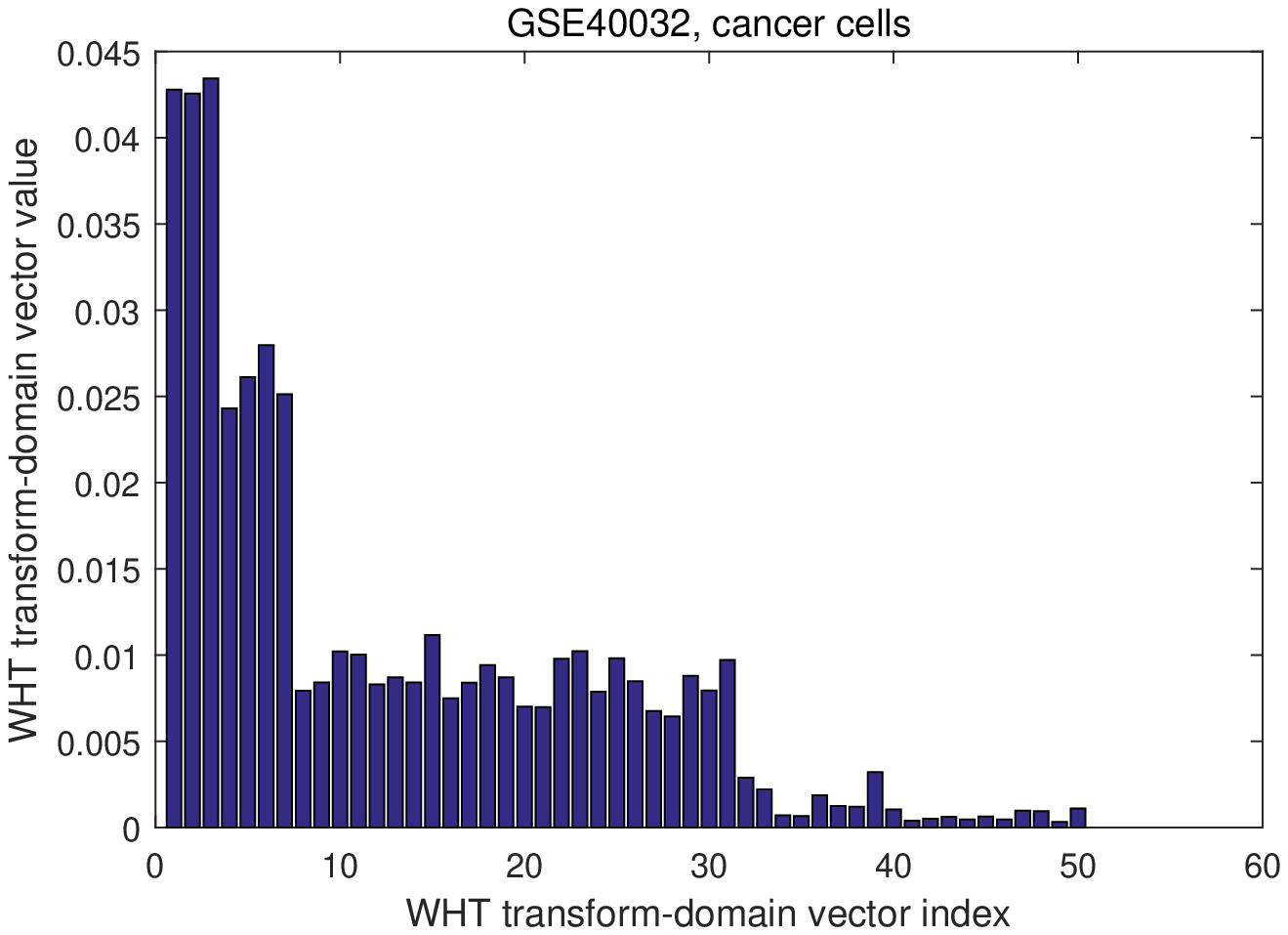}}\\
\subfigure[]{\includegraphics[width=9cm]{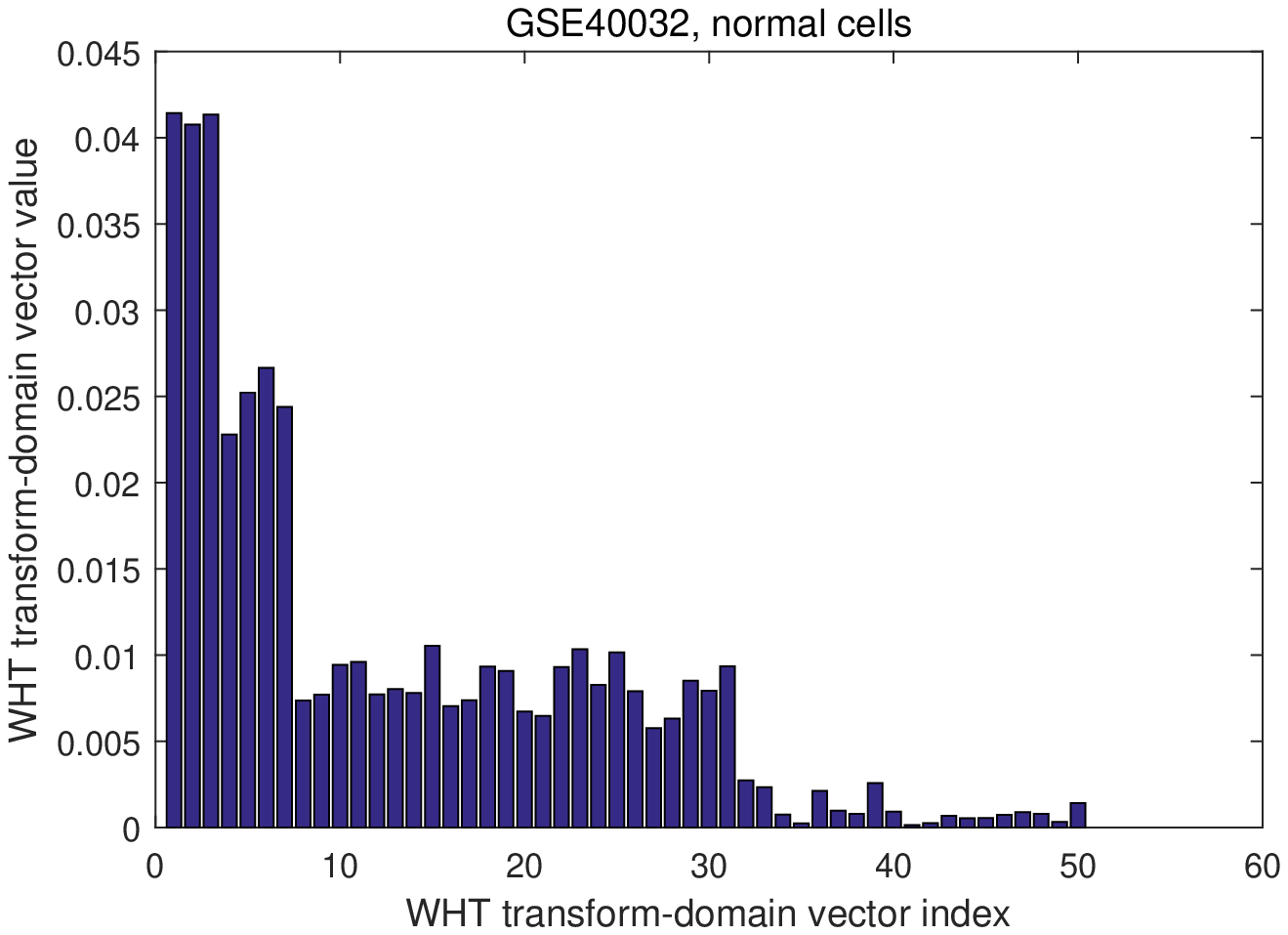}}
\end{varwidth}
\caption{The WHT transform vectors for NCBI datasets GSE63384 and GSE40032. Three-step values are observed for both cancer and normal cells in each dataset. This property is found to be a consistent trend.}
\label{transform_vectors}
\end{figure*}

\begin{figure*}
\centering
\begin{varwidth}{0.5\linewidth}  
\subfigure[]{\includegraphics[width=9cm]{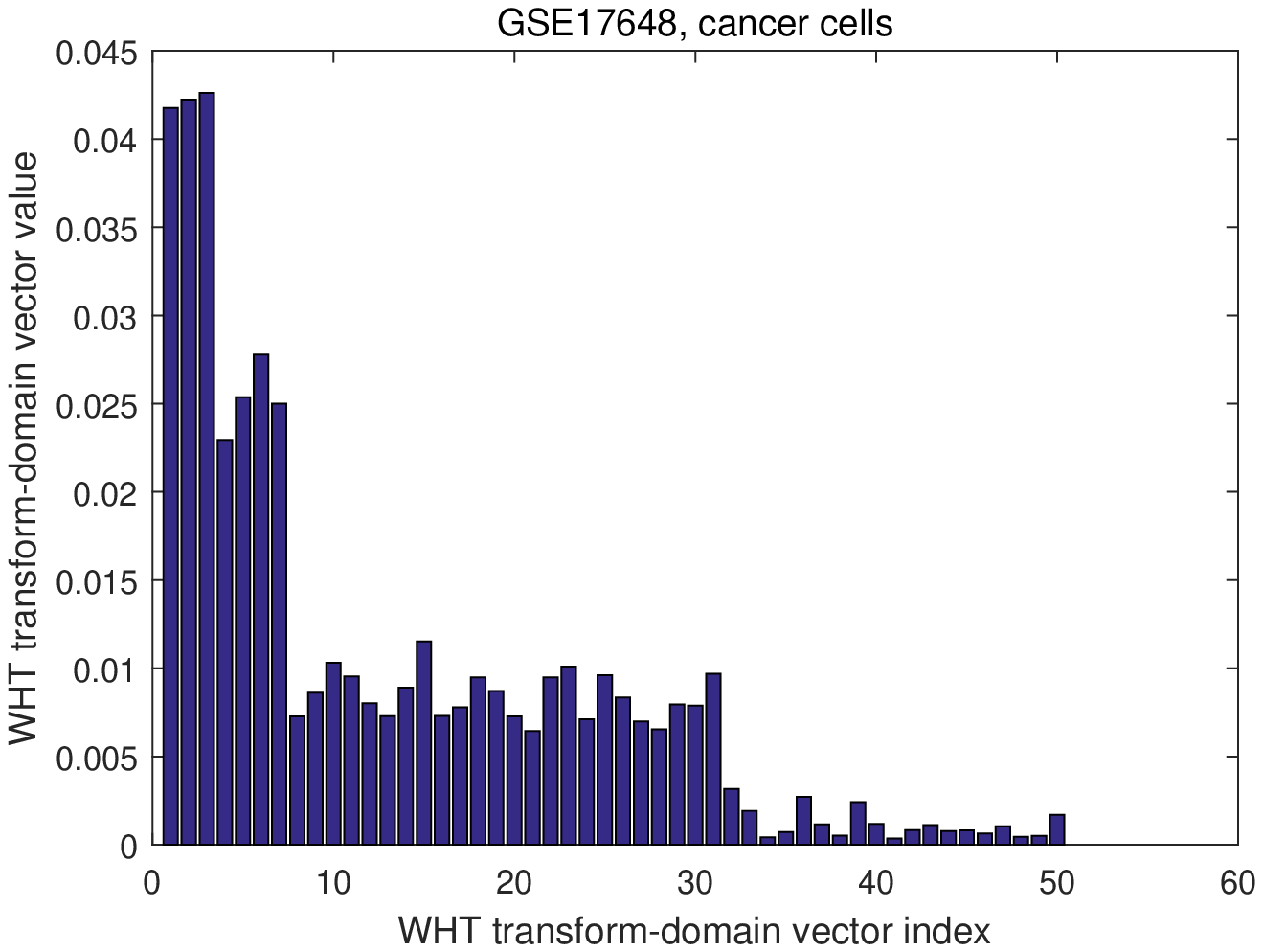}}\\
\subfigure[]{\includegraphics[width=9cm]{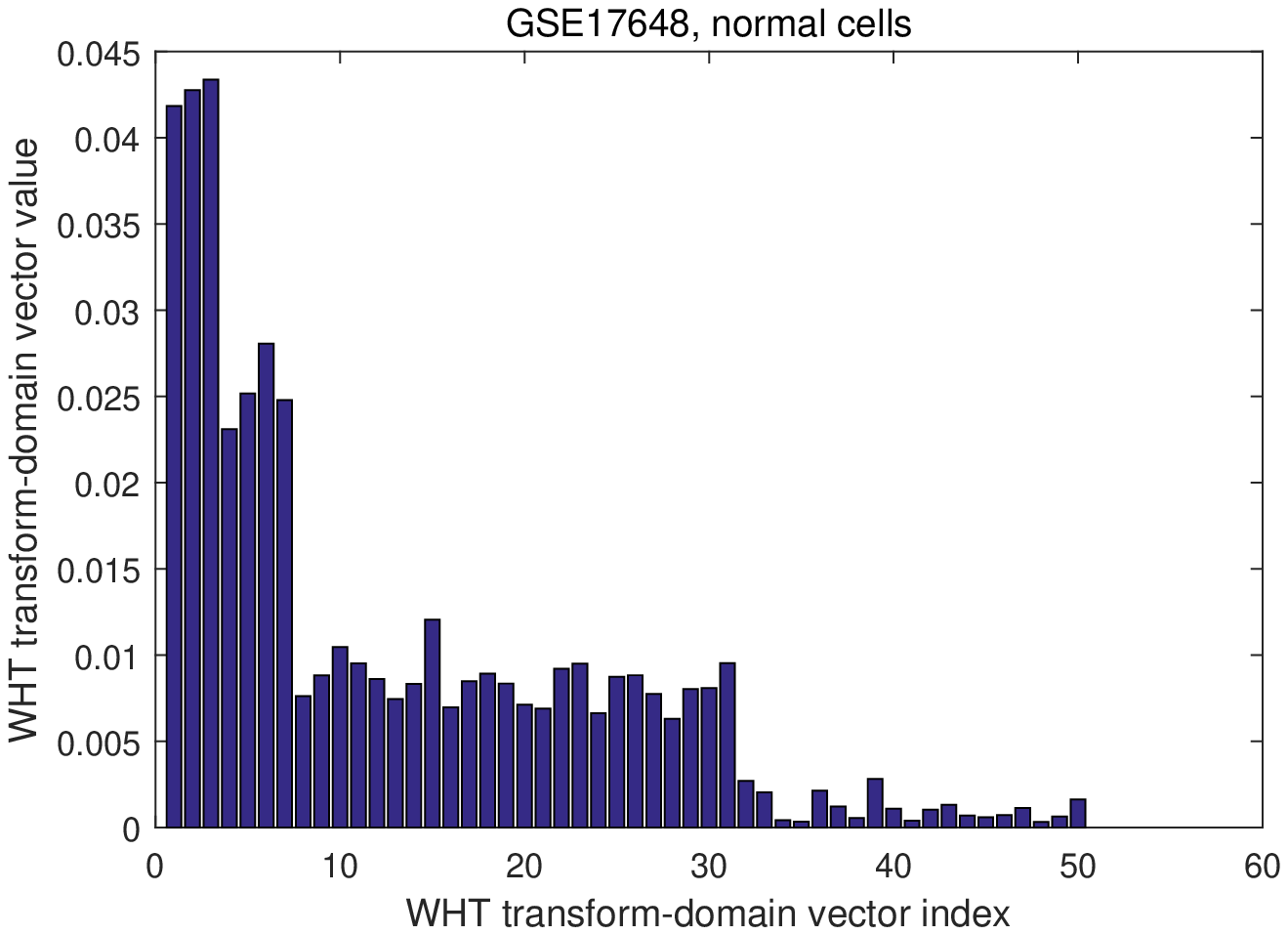}}
\end{varwidth}
\begin{varwidth}{0.5\linewidth}  
\subfigure[]{\includegraphics[width=9cm]{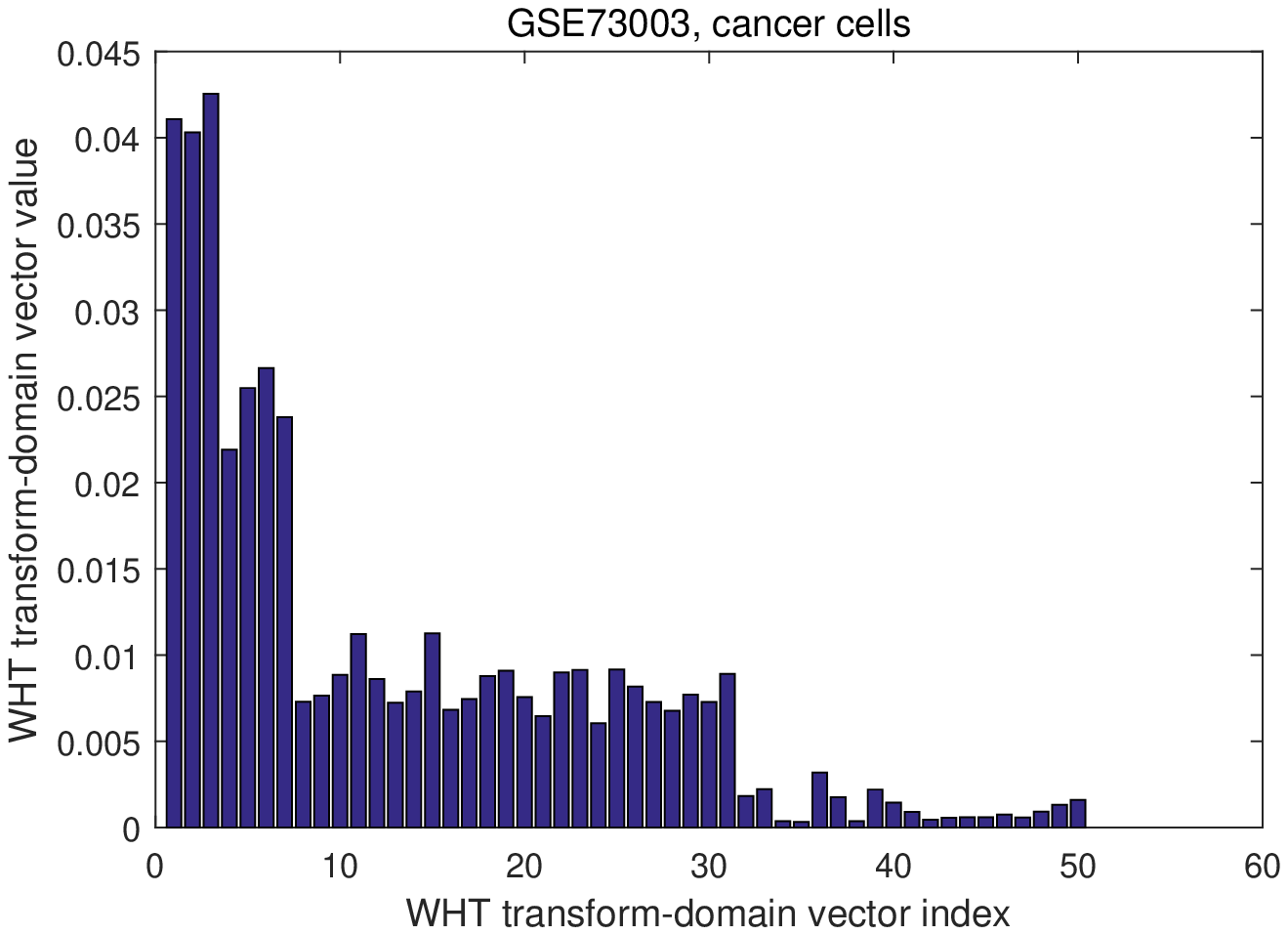}}\\
\subfigure[]{\includegraphics[width=9cm]{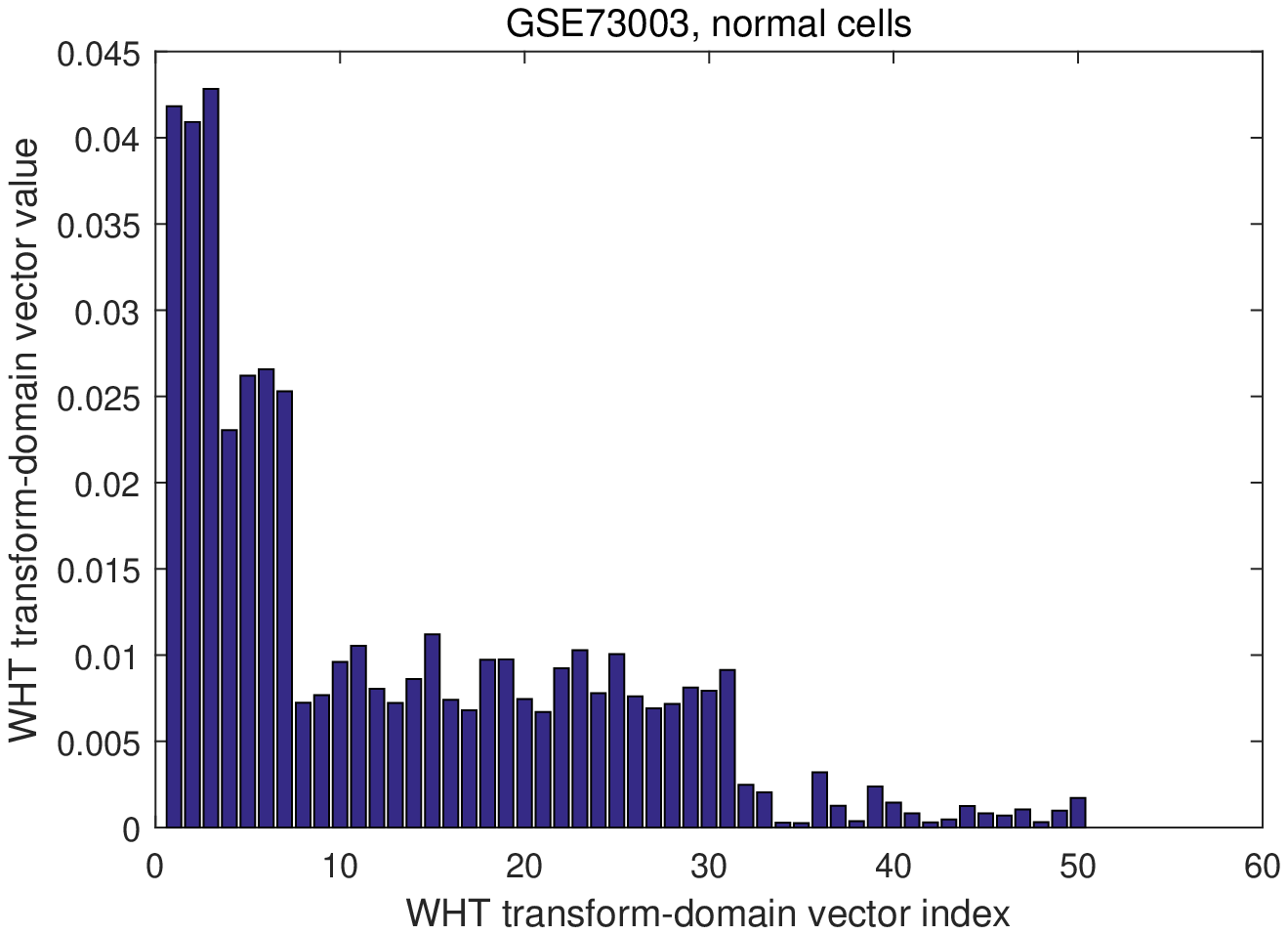}}
\end{varwidth}
\caption{The WHT transform vectors for NCBI datasets GSE17648 and GSE73003. Three-step values are observed for both cancer and normal cells in each dataset.}
\label{transform_vectors2}
\end{figure*}

\section{\bf{Methods}}\label{method}

In this work, the data in the analysis part is from the NCBI Gene Expression Omnibus~(GEO) datasets. GEO contains DNA methylation profiles of a large number of human diseases and tissue types, such as colorectal cancer cell, lung cancer cell, etc. Four datasets of cancer cells are selected and the beta values are the measured values for the CpG islands detected. The cell-classification results include the machine-learning and additional deep-learning results on the four datasets.

\textbf{Data acquisition procedure:}
The Human Methylation27K DNA Analysis BeadChip (Illumina) can measure methylation information at a single-base resolution for 27,578 CpG islands. After the bisulfite-conversion of the samples, the whole-genome amplification was performed with the input of bisulfite-converted DNA. The output was fragmented, purified and applied to the BeadChips by Illumina-supplied reagents and conditions. The array was stained and scanned after extension. The intensities of the unmethylated and methylated bead types were measured by the fluorescent signals.

\textbf{Extraction of CpG islands and computation of beta values:}
The DNA methylation signal intensities are extracted at the CpG islands measured by the DNA methylation Human Methylation27K BeadChip. The DNA methylation beta values are subsequently computed. The \emph{beta value}~\cite{Du10} is defined on each CpG island as $\beta(m)=\max(M(m),0)/(\max(M(m),0)+\max(U(m),0)+\alpha)$, where $m$ is the CpG island index, $M(m)$ is the methylated signal intensity, and $U(m)$ is the unmethylated signal intensity, and $\alpha$ is a normalization factor commonly chosen to be $100$. The methylated and unmethylated signal intensities are the fluorescent signal measurements. The beta value is in the range of $0$ to $1$, representing the level of methylation at particular CpG islands.

\textbf{Performing Walsh-Hadamard Transform~(WHT):}
We introduce now the low-dimension, three-step property of DNA methylation profile beta value vector after applying the Walsh-Hadamard Transform~(WHT). Firstly WHT is introduced; then the property after transformation is presented. The WHT is an orthogonal transform that is based on the Walsh matrix. The transform of length $K$ is defined by
\be
r(k) = \frac{1}{K}\sum\limits_{i = 0}^{K - 1} {a(i)W(k,i)},
\ee
where $a(i)$ is the $i$-th element in the data sequence of length $K$ before transformation, $W(k,i)$ is the $k$-th row and $i$-th column element of the Walsh matrix, and $r(k)$ is the $k$-th element of the vector after the transformation, with $k=0,1,...,K-1$. The Walsh matrix is defined recursively as,
\be
{\bf{W}}_n  = \left[ {\begin{array}{*{20}c}
   {{\bf{W}}_{n - 1} } & {{\bf{W}}_{n - 1} }  \\
   {{\bf{W}}_{n - 1} } & { - {\bf{W}}_{n - 1} }  \\
\end{array}} \right],
\ee
for $n \ge 1$, and ${\bf{W}}_0  = 1$. The $n$-th Walsh matrix can also be expressed by,
\be
{\bf{W}}_n  = {\bf{W}}_1  \otimes {\bf{W}}_{n - 1},
\ee
where $\otimes$ is the Kronecker product. WHT basically decomposes the original data sequence into a series of basis functions of Walsh matrix. The transform is also named Walsh Transform or Hadamard Transform in short. This transform has a fast implementation named Fast Walsh-Hadamard Transform~(FWHT), therefore it is suitable for the transformation of large-sized data vector.

\begin{figure}
\centering
\includegraphics[width=0.49\textwidth]{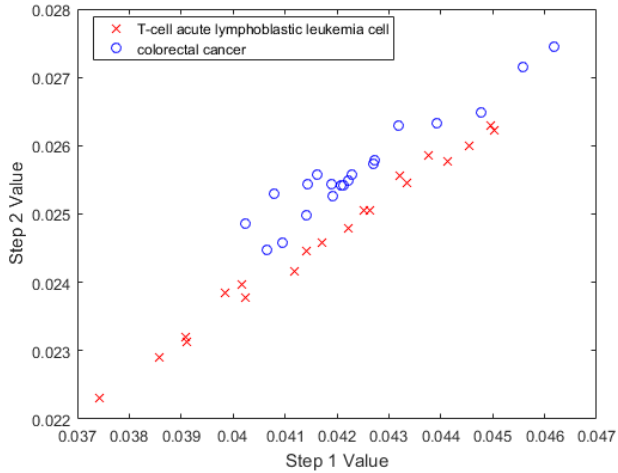}
\caption{Scatter plot of colorectal cancer cell~(GSE69954) and T-cell acute lymphoblastic leukemia cell~(GSE17648).}\label{fig_scattering}
\end{figure}

\textbf{Machine learning classification analysis:}
{
The disease and normal cell samples are processed with the pipeline followed by machine learning classification. The beta value vectors are firstly truncated to short sequence with varying length; then, the short sequences are analyzed by the machine learning classifiers on disease association. Classification accuracy results are generated for the k-fold cross-validation on several machine-learning classifiers and the preferred classifiers are selected for this classification task.
}

\begin{table*}[t]
 \caption{The classification accuracy comparisons of the whole original sequence classifiers and the WHT based classifiers. The k-fold cross-validation is applied to all the classifiers. In the table the value of $k$ in the k-fold cross-validation is noted. The classification accuracy with WHT is the averaged result of classification accuracies at three WHT feature space values equaling $83$, $89$ and $95$. }\label{table:accuracy}
\begin{center}
 \begin{tabular}{|p{3.5cm}|p{1.4cm}|p{1.4cm}|p{1.4cm}|p{1.4cm}|p{1.4cm}|p{1.4cm}|p{1.4cm}|p{1.4cm}|}
 \hline
 \textbf{Classifier/Validation Method} & \small{\textbf{GSE63384}} Whole Seq & \small{\textbf{GSE63384}} WHT & \small{\textbf{GSE17648}} Whole Seq & \small{\textbf{GSE17648}} WHT & \small{\textbf{GSE73003}} Whole Seq & \small{\textbf{GSE73003}} WHT & \small{\textbf{GSE40032}} Whole Seq & \small{\textbf{GSE40032}} WHT \\
 \hline\hline
 Ensemble Method of Boosting, k = sample size & 0.9286 & \bf{0.9381} & 0.9318 & \bf{0.9773} & 0.9500 & \bf{0.9250} & 0.9540 & \bf{0.9119} \\
 \hline
 Ensemble Method of Boosting, k = 3 & 0.9286 & \bf{0.9143} & 0.8864 & \bf{0.9621} & 0.8250 & \bf{0.8917} & 0.8966 &  \bf{0.9004} \\
 \hline
 Ensemble Method of Subspace, k = sample size & 0.6429 & 0.7476 & 0.8636 & 0.9848 & 0.8500 & 0.9250 & 0.7241 & 0.7356 \\
 \hline
 Ensemble Method of Subspace, k = 3 & 0.8000 & 0.7476 & 0.8636 & 0.9394 & 0.8500 & 0.9417 & 0.7356 &  0.7395 \\
 \hline
 Decision Tree, k = sample size & 0.8571 & 0.9429 & 0.9318 & 0.9318 & 0.9500 & 0.9000 & 0.8276 & 0.8563
 \\
 \hline
 Decision Tree, k = 3 & 0.8143 & 0.7714 & 0.9091 & 0.9015 & 0.7500  & 0.8583 & 0.8506 & 0.7714 \\
 \hline
 Support Vector Machine, k = sample size & 0.9143 & \bf{0.9143} & 0.9655 & \bf{0.9545} & 0.9500 & \bf{0.9667} & 0.8927 & \bf{0.9773} \\
  \hline
 Support Vector Machine, k = 3 & 0.9000 & \bf{0.9000} & 0.9773 & \bf{0.9773} & 0.9500 & \bf{0.9083} & 0.9655 & \bf{0.9119} \\
 \hline 
 Ensemble Method of Bagging, k = sample size & 0.9143 & 0.9190 & 1.0000 & 0.9924 & 0.9500 & 0.9500 & 0.9540 & 0.8736 \\
 \hline
 Ensemble Method of Bagging, k = 3 & 0.9286 & 0.8929 & 1.0000 & 0.9924 & 0.9500 & 0.9250 & 0.9425 & 0.8544 \\
 \hline
 k-Nearest Neighbors, k = sample size & 0.8000 & 0.7143 & 0.9091 & 1.0000  & 0.9000 & 0.9250 & 0.9195 & 0.8084  \\
 \hline
 k-Nearest Neighbors, k = 3 & 0.7714 & 0.7238 & 0.9091 & 0.9621 & 0.9250 & 0.8500 & 0.8966 & 0.7816 \\
 \hline
 Feedforward Neural Network, k = 10 & 0.8449 & 0.8640 & 0.9433 & 0.9961 & 0.8972 & 0.7132 & 0.9665 & 0.9351\\
  \hline
\end{tabular}
\vspace{-0.1in}
\end{center}
\end{table*}

\section{\bf{Results}}\label{results}

In this part, the three-step properties of the transform-domain vector are firstly reported. This discovered property is further applied to the classification of disease and normal cells. The classification accuracies and computation time are evaluated for various machine learning classifiers.

\textbf{Three-Step properties after WHT:}
We found that the beta value vector after WHT shows a \emph{three-step property}, and this property is consistently shown for the human tissue cells we have evaluated. To illustrate this property, WHT is performed on the beta value vectors of three NCBI GEO datasets and the three-step effect are plotted in Fig.~\ref{transform_vectors}, which depicts the WHT domain value vs. the domain index (from 2 to 100). These three datasets are beta value vectors measured by the HumanMethylation 27K bead chip. The length of the sequence before WHT transformation is approximately 27 thousand. It can be observed that the transform-domain vector has the three-step property with low dimension. The three-step property for 27K bead chip array is, there is \emph{one step} in the range of 2-nd to 4-th elements, which has the highest amplitude. The \emph{second step} is in the range of 5-th to 8-th elements with the amplitude lower than in the first step. The \emph{third step} is in the range of 9-th to 32-nd elements, and the amplitude is lower than in the second step. We have neglected the first element due to its very large amplitude compared with other samples. We have tested a large number of beta value vectors including the normal tissue cells and disease tissue cells and found that this property is consistently shown \emph{in all the human cell samples tested}. Based on this observation, we further propose disease and non-disease cell classification method.

The averaged step values and ranges of the steps of normal human tissue cell from the GEO datasets are given in Table~\ref{table_steps}. The step values in the table are computed from the normal tissue cells only in the selected GEO datasets. For each step value, the mean of one sample within the range of the step is firstly computed, then the step value is averaged over all the normal tissue cells. It is found that the three-step amplitude values are distinctively different for different human tissue cells; it is further found that the range of the step value only depends on the sequencing technique for obtaining the data. In Table~\ref{table_steps}, we have shown one dataset measured by Illumina HumanMethylation 450K beadchip. The transform-domain vector also shown three-step property, with different cut-off elements than 27K beadchip measurements. For 450K beadchip, the range of the first step is from the 2-nd to the 4-th element, and the range of the second step is from the 5-th to the 16-th element, and the range of the third step is from the 17-th to the 64-th element. We have neglected the first element because this value is much larger than the other elements and also varies greatly from sample to sample. Later classification algorithm also indicates that including the first element does not improve the classification accuracy. Thus, the first element is excluded in the step-range partition.

\begin{table*}[t]
 \caption{The computation time (in seconds) of the classification with the whole original sequence and the proposed WHT data. }\label{table:computation_time}
 \begin{center}
 \begin{tabular}{|p{3.3cm}|p{1.4cm}|p{1.4cm}|p{1.4cm}|p{1.4cm}|p{1.4cm}|p{1.4cm}|p{1.4cm}|p{1.4cm}|}
 \hline
 \textbf{Classifier} & \small{\textbf{GSE63384}} Whole Seq & \small{\textbf{GSE63384}} WHT & \small{\textbf{GSE17648}} Whole Seq & \small{\textbf{GSE17648}} WHT & \small{\textbf{GSE73003}} Whole Seq & \small{\textbf{GSE73003}} WHT & \small{\textbf{GSE40032}} Whole Seq & \small{\textbf{GSE40032}} WHT \\
 \hline\hline
 Ensemble Method of Boosting & 741.7 & \bf{57.12} & 355.06 & \bf{37.46} & 302.52 & \bf{34.37} & 1858.7 & \bf{71.45} \\
 \hline
Ensemble Method of Subspace & 1602.7 & 67.58 & 251.92 & 44.48 & 113.02 & 40.80 & 726.6 & 83.66  \\
 \hline
Decision Tree & 175.0 & 1.58 & 23.85 & 0.97 & 14.61 & 0.88 & 189.7 & 1.86 \\
 \hline
Support Vector Machine & 70.4 & \bf{2.25} & 44.00 & \bf{1.41} & 37.78 & \bf{1.30} & 95.20 & \bf{2.60} \\
 \hline
Ensemble Method of Bagging & 1302.6 & 65.08 & 203.48 & 42.21 & 133.35 & 38.49 & 5813.1 & 81.38 \\
 \hline
k-Nearest Neighbors & 88.5 & 1.63 & 6.93 & 1.01 & 5.46 & 0.93 & 213.3 & 1.80 \\
 \hline
Feedforward Neural Network & 229.2 & 0.86 & 149.82 & 0.93 & 116.65 & 0.73 & 1963.0 & 1.68\\
 \hline
\end{tabular}
\vspace{-0.1in}
\end{center}
\end{table*}

{
The three-step property of the WHT transform-domain vector is related to the distribution of the beta values controlled by the chromatin state of human cells. The chromatin state is characterized by the modifications on the eight histone proteins, the methylation on DNA sequence, and the chromatin-associated proteins in a particular genome sequence. The chromatin state can be modelled by Hidden Markov Model~(HMM)~\cite{Ernst17, Marco17}, where the hidden states of the HMM are the states of chromatin; and the observed outputs of the HMM are the measured DNA methylation and histone modification. There are different chromatin states for different types of human cells~\cite{Ernst11}. The DNA methylation intensity measurements at the CpG islands are controlled by the chromatin state. The beta value distribution is determined by the DNA methylation intensity measurements at the CpG islands. The three-step property is determined by the distribution of the beta values. Thus the chromatin state of human cell is the major reason of the three-step property in the transform domain vector of the DNA methylation beta values of human cells. The human cells have chromatin states producing the DNA methylation intensity measurements that create the three-step property of the beta values after WHT transformation.
}

 
It is informative to note that, the transform-domain vector can be applied to classify different types of cells. In Fig.~\ref{fig_scattering} the scatter plot of two different types of human tissue cells, the colorectal cancer cell~(GSE69954) and the T-cell acute lymphoblastic leukemia cell~(GSE17648). The mean of the first and second steps are the classification features for classifying these two types of cells in the disease condition. It can be observed from the scatter plot that the samples of the two cells are far apart; thus, a linear classifier can easily distinguish these two type of cells. Therefore, the mean of the step values can be the feature space for the classification of different human tissue cells. By testing, we find that the classification accuracy of the cell types by the Support Vector Machine in GSE69954 and GSE17648 can achieve $100\%$ accuracy for the two types of issues. This is because, for different types of cells the scatter plot samples are well separated in the plane as indicated in Fig.~\ref{fig_scattering}. Due to this fact, the cell type classification is not the focus on this work. 

The disease cell classification is a more difficult problem than cell type classification due to the subtle difference in the DNA methylation profile between disease cells and normal cells. The scatter plots of disease and non-disease cells of the same type of tissue have significant overlap. The major classification problem becomes how to classify the disease and non-disease cells for the same tissue type. It is thus necessary to test the machine learning classification algorithms on disease and non-disease cells of the same tissue type. In the following section, we show the results to classify the disease and non-disease cells by the proposed method.

\textbf{Disease cell classifications:}
{
In this part, the disease cell classification is done for WHT and the whole original sequence. The numbers of disease and normal cells in the four types of NCBI datasets are given in Table~\ref{table_datasets}. Since the samples in the analysis are from NCBI datasets, the procedure to obtain these samples is following the respective groups that perform the data collection. Since all datasets are collected on the HumanMethylation 27K beadchip platform, the procedures are the same for the four datasets applied in this research. The experimental procedure to obtain these datasets is explained in Sect.~\ref{method}. The machine learning classification methods we considered are: Ensemble Method of Boosting, Ensemble Method of Bagging, Ensemble Method of Random Subspace, Decision Tree, Support Vector Machine, k-Nearest Neighbors, and Feedforward Neural Network. For the Ensemble Method of Boosting, the algorithm of LogitBoost is applied. For WHT data, these machine learning classifiers are applied on the truncated transform-domain data. Based on the discussions in previous sections, the proposed WHT-based transformation can reduce the length of the beta value vector to less than $100$ for both Illumina HumanMethylation 27K beadchip and Illumina HumanMethylation 450K beadchip. Note that, the latent low-dimensional feature of DNA methylation is discovered by WHT, therefore in the article all the machine learning algorithms are compared with the WHT assumption. The comparisons with other latent feature methods are out of the scope of this work.

The classification accuracy performance is evaluated with k-fold cross-validation. There are two k values selected for the performance evaluation---(i)~$k$ = sample size: one sample is for each validation and the remain samples are for training. The training and validation are done for the times equaling to the sample size. The final classification accuracy performance result is obtained by averaging these intermediate results. (ii)~$k=3$: the datasets are partitioned into three folds, two folds for the training and one fold for the validation. The training and validation are done three times with three setups, and the final classification accuracy results are obtained by averaging the three results. The performance results of the WHT data and the original whole sequence data are obtained in Table~\ref{table:accuracy}. The computation running time results are given in Table~\ref{table:computation_time}. The running time include the time of WHT computation, k-fold cross-validation training and inference of the classifiers evaluated.

From Table~\ref{table:accuracy}, we can observe that the Ensemble Method of Boosting and Support Vector Machine achieve comparably high accuracy performance among all the classifiers evaluated. With Ensemble Method of Boosting and Support Vector Machine, the WHT has the close performance to the whole sequence, for both k-fold cross-validation options including $k$ equaling to the sample size and $k$ equal to $3$. For the Ensemble Method of Boosting and $k$ equal to sample size, with dataset GSE63384 the classification accuracy is $0.9381$ and $0.9286$ for WHT and whole sequence; with dataset GSE40032, the classification accuracy is $0.9119$ and $0.9540$ for WHT and whole sequence; with dataset GSE17648 the accuracy is $0.9773$ and $0.9318$ for WHT and whole sequence; with dataset GSE73003 the accuracy is $0.9250$ and $0.9500$ for WHT and whole sequence. For the Ensemble Method of Subspace, the proposal has performed better than the whole sequence classification for all the four datasets. For the classifier of Decision Tree, the proposed WHT method performs slightly better for GSE63384 and GSE40032 to the whole sequence classification. Similarly, with Support Vector Machine, WHT also achieves close performance to the whole sequence for both k-fold cross-validation options. 

It can also be observed that for the classifiers of Ensemble Method of Boosting and Support Vector Machine, the overall performance of WHT is at the same level to the performance of the whole sequence---the classification accuracy of WHT is slightly higher than with the whole sequence in some cases, and slightly below in some other cases. The classification accuracies of WHT are in bold in Table~\ref{table:accuracy} for the two selected classifiers of Ensemble Method of Boosting and Support Vector Machine. It is noted that for the Ensemble Method of Bagging there is one dataset GSE40032 that has performance discrepancy between WHT and whole sequence, therefore Ensemble Method of Bagging is not selected as the preferred classifier. It can also be observed that the following classifiers do not produce high classification accuracy results, therefore, are not selected: Ensemble Method of Subspace, Decision Tree, k-Nearest Neighbors, and Feedforward Neural Network. Note that, the Feedforward Neural Network adopts the basic structure of one input layer, and one hidden layer and one output layer.

From the running time results in Table~\ref{table:computation_time}, it can be observed that the proposed WHT based classifiers achieve one order of magnitude reduction in computation time compared with the whole sequence classification. For the Ensemble Method of Boosting, the proposed WHT based method achieves the following folds of running time speed improvement compared with the whole sequence classification: $13$ folds for GSE63384, $9.5$ folds for GSE17648, $8.8$ folds for GSE73003, and $26$ folds for GSE40032. The run time speed up folds observed for the classifier of Support Vector Machine are: $31$ folds for GSE63384, $31$ folds for GSE17648, $29$ folds for GSE73003, and $36$ folds for GSE40032. Together with the classification accuracy performance in Table~\ref{table:accuracy}, it is identified that the Ensemble Method of Boosting and Support Vector Machine with WHT achieve comparable performance compared with the whole sequence while having significant advantages of reduced computational time. All the training is done on the same qual-core desktop computer so that the computational time durations of different machine learning algorithms are compared under the same hardware configuration. Note that the data in Table~\ref{table:accuracy} is obtained by averaging the last three WHT feature space vector length values equal $83$, $89$, and $95$. The reason we have selected the last three points to produce the results in Table~\ref{table:accuracy} is that there are fluctuations while varying the feature space vector length, and the last three points are the most stable results. 

}

\section{\bf{Conclusion}}\label{conclusion}

We presented a novel human disease cell-classification approach based on the three-step property of the DNA methylation profile with the Walsh-Hadamard Transform~(WHT). The feature space vector is chosen as the WHT transform domain vector, and classification accuracy results are presented for competing machine-learning classifiers. We found that our proposal can achieve significant running time reduction while maintaining comparable performance compared with the original sequence classification for the identified classifiers include the Ensemble Method of Boosting and Support Vector Machine. Our proposal has valueable practical applications to the disease human cell classification tasks that require expedited, real-time computation. This framework has the perspectives to be applied to human cell classifications of other disease types and be generalized to other types of epigenome and genome datasets.


\begin{small}
\bibliographystyle{ieeetr}
\bibliography{Reference_DNA_methylation}
\end{small}

\end{document}